\documentclass[preprint,showpacs,preprintnumbers,amsmath,amssymb]{revtex4}

\usepackage{amssymb}

\newtheorem{prop}{Proposition}[section]

\newcommand\RR{\mathbb{R}}

\begin{document}
\def\eqn#1{Eq.$\,$#1}
\def\mb#1{\setbox0=\hbox{$#1$}\kern-.025em\copy0\kern-\wd0
\kern-0.05em\copy0\kern-\wd0\kern-.025em\raise.0233em\box0}
\preprint{}

\title{Chapman-Enskog derivation of the generalized \\
Smoluchowski equation}
\author{Pierre-Henri Chavanis$^{1}$, Philippe Lauren\c cot$^{2}$ and
  Mohammed Lemou$^{2}$} 
\affiliation{(1) Laboratoire de Physique Th\'eorique,  Universit\'e
Paul Sabatier, 118 route de Narbonne, 
31062 Toulouse Cedex 4, France\\
(2) Math\'ematiques pour l'Industrie et la Physique, CNRS UMR~5640,
Universit\'e Paul Sabatier
118 route de Narbonne, 31062 Toulouse Cedex 4, France}

\begin{abstract}

We use the Chapman-Enskog method to derive the Smoluchowski equation
from the Kramers equation in a high friction limit. We consider two
main extensions of this problem: we take into account a uniform
rotation of the background medium and we consider a generalized class
of Kramers equations associated with generalized free energy
functionals. We mention applications of these results to systems with
long-range interactions (self-gravitating systems, 2D vortices,
bacterial populations,...). In that case, the Smoluchowski equation is
non-local. In the limit of short-range interactions, it reduces to a
generalized form of the Cahn-Hilliard equation. These equations are
associated with an effective generalized thermodynamical
formalism. 

\vskip1cm
\noindent Corresponding author: P.H. Chavanis; e-mail: chavanis@irsamc.ups-tlse.fr; Tel: +33-5-61558231; Fax: +33-5-61556065

\end{abstract}

\maketitle

\newpage

\section{Introduction}
\label{sec_introduction}

The Kramers (or Klein-Kramers) equation \cite{kramers,klein} and the
associated Smoluchowski equation \cite{smoluchowski} are widely
discussed in the physics of Brownian motion and stochastic processes.
Initially, the Kramers equation was introduced to describe the
statistical evolution of a gas of Brownian particles in a potential
well. It has been applied in different domains of physics such as
colloidal suspensions, stellar systems, chemical reactions rate
theory, nuclear dynamics, just to mention a few \cite{chandra}. The
Kramers equation forms a particular class of Fokker-Planck equations
with a constant diffusion coefficient and a linear friction
\cite{risken}. In the large time limit, the velocity distribution of the
particles is close to the Maxwellian distribution and the evolution of
the spatial density is described by the Smoluchowski equation. The
passage from Kramers to Smoluchowski equations has been the subject of
many papers. In the early days, this passage was informal and
qualitative \cite{kramers,chandra}. The derivation of the Smoluchowski
equation from the Kramers equation is now a classical problem in many
textbooks of statistical mechanics.  In his monograph, van Kampen
presents an expansion procedure similar to the Hilbert perturbation
scheme \cite{kampen}.

In this paper, we propose another derivation of the Smoluchowski
equation based on the Chapman-Enskog expansion. Usually, the
Chapman-Enskog perturbation scheme is used to derive the
hydrodynamical equations (Euler, Navier-Stokes,...) from the kinetic
Boltzmann equation
\cite{huang}. In this context, the small parameter is played by the
typical collision time $\tau$. The Chapman-Enskog expansion allows to
determine the expression of the diffusion coefficients (viscosity,
heat conductivity,...)  in the fluid equations. In the context of the
Kramers equation, the small parameter is played by the inverse of the
friction coefficient $\xi$. We can therefore expand the distribution
function in powers of $\xi^{-1}$. To first order in this expansion, we
obtain the Smoluchowski equation. The diffusion coefficient is given
by the Einstein relation $D=k_{B}T/\xi m$ where $T$ is the temperature.

We shall consider two main generalizations of this problem. First, we
consider the situation in which the ambiant medium is uniformly
rotating so that the velocity which appears in the Kramers equation is
the {\it relative} velocity with respect to the background flow.
Inertial forces appear in the rotating frame and the expression of the
Smoluchowski equation is modified. Secondly, we consider a generalized
class of Kramers equations introduced recently by Kaniadakis
\cite{kaniadakis}, Frank \cite{frank}
and Chavanis \cite{gfp,gkt} (see also an early version in
\cite{csr}). These equations share the same properties as the ordinary
Kramers equation but they are associated with a more general form of
free energy functional.  This generalization encompasses the case of
quantum statistics (fermions and bosons) and it can also find
applications in the physics of complex media. In that context, they
extend the nonlinear Fokker-Planck equation
\cite{plastino,buckman} associated with the Tsallis entropy \cite{tsallis}.  
We derive here the generalized Smoluchowski equation from the
generalized Kramers equation by using a systematic procedure. The
diffusion coefficient is given by a generalized Einstein relation
which depends on a function $\phi(x)$ implicitely determined by a
second order differential equation.

The paper is organized as follows. In Sec.~\ref{sec_gks}, we recall
the main properties of the (generalized) Kramers and Smoluchowski
equations. In Sec.~\ref{sec_ce}, we apply the Chapman-Enskog expansion
to the generalized Kramers equation. We account for a solid rotation
of the background medium and calculate the new terms that appear in
the Smoluchowski equation due to rotation. In Sec. \ref{sec_mom}, we
derive a variant of the generalized Smoluchowski equation by using a
moment method.  Finally, in Sec. \ref{sec_lr}, we shortly discuss the
application of these results to the case of systems with long-range
interactions (self-gravitating systems, 2D vortices, bacterial
populations,...). In that case, the Smoluchowski equation becomes
non-local and possesses a rich mathematical and physical structure
\cite{gfp}. We determine the Lyapunov functional associated with the
generalized Smoluchowski equation and interpret this functional as a
free energy. In the limit of short-range interactions, this functional
reduces to the Landau free energy and the non-local generalized
Smoluchowski equation reduces to a generalized form of the
Cahn-Hilliard equation. In Appendix~\ref{sec_bgk}, we consider the
case of an isotropic BGK operator and in Appendix~\ref{sec_time}, we
consider a wider class of Kramers equations with time varying Lagrange
multipliers accounting for integral constraints (energy, angular
momentum and impulse conservation).

\section{Generalized Kramers and Smoluchowski equations}
\label{sec_gks}

It has been noted recently that standard kinetic equations (Boltzmann,
Landau, Kramers, Smoluchowski,...) could be generalized so that they
increase a larger class of functionals than the Boltzmann entropy (or
Boltzmann free energy) \cite{gkt}. Such
generalized kinetic equations appear when the transition probabilities
have an expression different from the one we would naively expect
\cite{kaniadakis} or when the system is described by a stochastic
process involving a multiplicative noise depending on the density
\cite{borland,gfp}. These generalized kinetic equations
encompass the class of quantum kinetic equations (with exclusion or
inclusion principle) already discussed in the literature. They can
also be interpreted as ``effective equations'' trying to take into
account ``hidden constraints'' in the physics of complex media \cite{gkt}.

The generalized Kramers equation can be written in the form \cite{gfp}:
\begin{equation}
\label{gsk1}
{\partial f\over\partial t}+{\bf v}\cdot{\partial f\over\partial {\bf
    r}}-\nabla\Phi\cdot{\partial f\over\partial {\bf
    v}}={\partial\over\partial {\bf v}}\cdot\biggl\lbrace
D\biggl\lbrack f C''(f){\partial f\over\partial {\bf v}}+\beta f {\bf
  v}\biggr\rbrack\biggr\rbrace,
\end{equation}
where $f=f({\bf r},{\bf v},t)$ is the distribution function, $C$ is a
convex function (i.e. $C''>0$), $\beta=1/T$ is the inverse temperature
and $\Phi({\bf r})$ is an external potential. Since the temperature
$T$ is fixed, the above equation describes a canonical situation. We
introduce the functional
\begin{equation}
\label{gsk2}
F[f]={1\over 2} \int f v^{2}d^{3}{\bf r}d^{3}{\bf v}+\int \rho
\Phi d^{3}{\bf r}+T \int C(f)d^{3}{\bf r}d^{3}{\bf v},
\end{equation}
where $\rho=\int f d^{3}{\bf v}$ is the spatial density.  This
functional can be interpreted as a free energy $F=E-TS$, where $S$ is
a ``generalized entropy'' and $E=K+W$ is the energy including a
kinetic term and a potential term. When $S[f]$ is the Boltzmann
entropy, Eq. (\ref{gsk1}) reduces to the ordinary Kramers
equation. More generally, it is straightforward to check that $F[f]$
plays the role of a Lyapunov functional satisfying $\dot F\le 0$. This
is similar to a canonical version of the H-theorem. Finally, a
stationary solution of Eq. (\ref{gsk1}) is determined by the condition
$\partial f/\partial t=0$, implying $\dot F=0$, and yielding
\begin{equation}
\label{gsk3}
C'(f_{eq})=-\beta \biggl ({v^{2}\over 2}+\Phi({\bf r}) \biggr )-\alpha.
\end{equation}
The equilibrium distribution $f_{eq}({\bf r},{\bf v})$ minimizes the
free energy $F[f]$ at fixed mass and temperature. Similarly, the
generalized Kramers equation (\ref{gsk1}) maximizes the rate of
dissipation of free energy $\dot F$ at fixed mass and temperature. We
refer to Ref. \cite{gfp} for a more detailed discussion of these
results.

For large times $t\gg \xi^{-1}$, where $\xi=D\beta$ is the friction
coefficient, the distribution function is close to the {\it isotropic}
distribution function $f({\bf r},{\bf v},t)$ determined  by the
relation
\begin{equation}
\label{gsk4}
C'(f)=-\beta \biggl \lbrack {v^{2}\over 2}+\lambda({\bf r},t)\biggr \rbrack.
\end{equation}
This distribution function cancels out the diffusion current in the
generalized Kramers equation (\ref{gsk1}). The pressure $p={1\over
3}\int f v^{2} d^{3}{\bf v}=p(\lambda)$ and the density $\rho=\int f d^{3}{\bf
v}=\rho(\lambda)$ are related to each other by a barotropic equation of state
$p=p(\rho)$ which is entirely specified by the function $C(f)$. Taking
the moments of the generalized Kramers equation (\ref{gsk1}), it is
shown in \cite{gfp} that the time evolution of the density $\rho({\bf
r},t)$ is governed by the generalized Smoluchowski equation
\begin{equation}
\label{gsk5}
{\partial\rho\over\partial t}=\nabla\cdot\biggl \lbrack
{1\over\xi}(\nabla p+\rho\nabla\Phi)\biggr\rbrack.
\end{equation}
This equation was initially obtained in \cite{csr} from a special form
of Kramers equation associated with the Fermi-Dirac entropy.  The generalized Smoluchowski equation decreases the functional
\begin{equation}
\label{gsk6}
F[\rho]= \int \rho\int_{0}^{\rho}{p(\rho')\over\rho^{'2}}d\rho'd^{3}{\bf r} + \int\rho\Phi d^{3}{\bf r},
\end{equation}
which is the simplified form of free energy (\ref{gsk2}) obtained by using the
isotropic distribution function (\ref{gsk4}) to express $F[f]$ as a functional
of $\rho$ \cite{gfp}.  Thus, $\dot F\le 0$, as can be checked by a direct
calculation. At equilibrium, $\dot F=0$ and we obtain the condition of
hydrostatic equilibrium
\begin{equation}
\label{gsk7}
\nabla p+\rho\nabla\Phi={\bf 0}.
\end{equation}
This relation also directly results from Eq. (\ref{gsk3}).  The
object of this paper is to present a formal derivation of the
generalized Smoluchowski equation (\ref{gsk5}) from the generalized
Kramers equation (\ref{gsk1}), using a Chapman-Enskog expansion.

\section{Chapman-Enskog expansion of the generalized Kramers equation}
\label{sec_ce}

We write the generalized Kramers equation in the form
\begin{equation}
\label{ce1}
{\partial f\over\partial t}+{\bf v}\cdot{\partial f\over\partial {\bf
    r}}-\nabla\Phi\cdot{\partial f\over\partial {\bf
    v}}={\partial\over\partial {\bf v}}\cdot\biggl\lbrace D\biggl\lbrack
h(f){\partial f\over\partial {\bf v}}+\beta g(f) {\bf
  w}\biggr\rbrack\biggr\rbrace,
\end{equation}
where $g$ and $h$ are arbitrary (positive) functions \cite{gkt}. We
assume that our Brownian particles evolve in a background that is
rotating and translating uniformly, so that ${\bf w}={\bf v}-{\bf u}$
denotes the {\it relative} velocity of the particles with respect to
the background flow with ${\bf u} ={\mb\Omega}{\times}{\bf r}+{\bf
  U}$. The ``generalized entropy'' associated with this equation can be
written
\begin{equation}
\label{ce2}
S[f]=-\int C(f)d^{3}{\bf r}d^{3}{\bf v},
\end{equation}
where the function $C$ is determined by the relation
\begin{equation}
\label{ce3}
C''(f)={h(f)\over g(f)}.
\end{equation}

We consider the high friction limit $\xi=D\beta\rightarrow +\infty$,
where $\beta=O(1)$. We rewrite Eq. (\ref{ce1}) in the form
\begin{equation}
\label{ce4}
{\partial f\over\partial t}+L f={1\over\epsilon}Q(f),
\end{equation}
where $\epsilon=D^{-1}$ is a small parameter, $L={\bf v}\cdot
{\partial/\partial {\bf r}} - \nabla\Phi\cdot{\partial/\partial {\bf v}}$
is the advective operator and $Q(f)$ is the collision operator
\begin{equation}
\label{ce5}
Q(f)={\partial\over\partial {\bf v}}\cdot\biggl\lbrack h(f){\partial
  f\over\partial {\bf v}}+\beta g(f) {\bf w}\biggr\rbrack.
\end{equation}

Our aim is now to derive an equation for $\rho({\bf r},t)$ to second
order with respect to $\epsilon$. For that purpose, we follow a
Chapman-Enskog procedure and employ a formalism in the spirit of
\cite{DLP}. More precisely, we look for a
solution $f$ to (\ref{ce4}) of the form $f=f_{0}+\epsilon f_{1}$ where
$f_0$ is the equilibrium distribution function with the same density as
$f$, that is, $Q(f_0)=0$ and
\begin{equation}
\label{ce8}
\int f_{0}({\bf r},{\bf v},t)d^{3}{\bf v} = \rho({\bf r},t).
\end{equation}
Substituting this expression in Eq. (\ref{ce4}), we get
\begin{eqnarray}
\label{ce5.5}
\biggl ({\partial \over\partial t}+L\biggr )(f_{0}+\epsilon
f_{1})&=&{1\over\epsilon}Q(f_{0}+\epsilon f_{1}),\\
\label{ce6}
&=&DQ(f_{0})f_{1}+O(\epsilon),
\end{eqnarray}
where $DQ(f_{0})$ is the linearized collision operator given by
\begin{equation}
\label{ce15}
DQ(f_{0})f_{1}={\partial\over\partial {\bf v}}\cdot\biggl\lbrack
h'(f_{0})f_{1}{\partial f_{0}\over\partial {\bf v}}+h(f_{0}){\partial
  f_{1}\over\partial {\bf v}}+\beta g'(f_{0})f_{1}{\bf w}\biggr
\rbrack.
\end{equation}
It then follows from Eq. (\ref{ce6}) that $f_{1}$ satisfies
\begin{equation}
\label{ce7}
DQ(f_{0})f_{1}=\biggl ({\partial \over\partial t}+L\biggr )f_{0}+O(\epsilon).
\end{equation}
Next, integrating Eq. (\ref{ce1}) over the velocities, we obtain
\begin{equation}
\label{ce9}
{\partial\rho\over\partial t}+\nabla\cdot\biggl (\int f_{0}{\bf v}d^{3}{\bf
  v}\biggr )=-\epsilon\nabla\cdot\biggl (\int f_{1}{\bf v}d^{3}{\bf
  v}\biggr ).
\end{equation}

Our aim now is to express $f_{0}$ and $f_{1}$ in terms of $\rho$, so
as to deduce from (\ref{ce9}) a closed equation for $\rho$. First, we
note that $Q(f_{0})=0$ is equivalent to the condition
\begin{equation}
\label{ce10} C'(f_{0})=-\beta \biggl\lbrack {w^{2}\over 2}+
\lambda({\bf r},t)\biggr\rbrack.
\end{equation}
This relation, supplemented by  Eq. (\ref{ce8}), completely specifies
$f_{0}$.

We next determine $f_1$. For that purpose, in view of (\ref{ce7}), we
need to investigate the invertibility properties of $DQ(f_0)$. This is
the subject of the following result.

\begin{prop}\label{prlin}
The operator $DQ(f_0)$ is a non-positive self-adjoint and closed
operator on the Hilbert space $\mathcal{H}=L^2(\RR^3,C''(f_0)d^{3}{\bf
  v})$. The null space $\mathcal{N}$ of $DQ(f_0)$ is given by
\begin{equation}
\label{nullspace}
\mathcal{N} = \RR {1\over C''(f_0)} = \RR {g(f_0)\over h(f_0)}=\biggl\lbrace \lambda {g(f_{0})\over h(f_{0})},\lambda\in\RR\biggr \rbrace,
\end{equation}
and the range of $DQ(f_0)$ is the orthogonal space to $\mathcal{N}$,
denoted by $\mathcal{N}^\bot$.
\end{prop}

\noindent\textsl{Proof.} We observe that the linearized collision operator
takes the simpler form
\begin{equation}
\label{ce17}
DQ(f_{0})F={\partial\over\partial {\bf v}}\cdot\biggl\lbrack
g(f_{0}){\partial  \over\partial {\bf v}}\left( C''(f_0) F \right)
\biggr \rbrack.
\end{equation}
Consequently, for $F\in\mathcal{H}$ and $G\in\mathcal{H}$, we have
$$
\langle DQ(f_0)F,G\rangle_{\mathcal{H}} = - \int g(f_0)
{\partial\over\partial {\bf v}}\left( C''(f_0) F \right)\cdot
{\partial\over\partial {\bf v}}\left( C''(f_0) G \right)d^3{\bf v},
$$
from which we readily deduce that $DQ(f_0)$ is non-positive and
self-adjoint in $\mathcal{H}$. Also, its null space is given by
Eq. (\ref{nullspace}). The closedness of $DQ(f_0)$ finally follows
by classical arguments. \hfill $\square$

\medskip

We are now in a position to determine $f_1$. Let $\Pi$ be the
orthogonal projection in $\mathcal{H}$ on $\mathcal{N}$ and denote by
$DQ(f_0)^{-1}$ the inverse of the restriction of $DQ(f_0)$ to
$\mathcal{N}^\bot$. Then, applying successively $Id - \Pi$ and
$DQ(f_0)^{-1}$ to (\ref{ce7}), we obtain
\begin{equation}
\label{f1}
f_1 = DQ(f_0)^{-1}\ (Id-\Pi)\ \biggl ({\partial \over\partial
  t}+L\biggr )f_{0} + O(\epsilon).
\end{equation}
Now, noting that
\begin{equation}
\label{ce11}
C''(f_{0}){\partial f_{0}\over\partial {\bf v}}=-\beta {\bf w},
\end{equation}
\begin{equation}
\label{ce12}
C''(f_{0}){\partial f_{0}\over\partial {\bf r}}=-\beta
\nabla\lambda+\beta \left( \nabla {\bf u} \right)^T {\bf w},
\end{equation}
and
\begin{equation}
\label{ce13}
C''(f_{0}){\partial f_{0}\over\partial t}=-\beta
{\partial\lambda\over\partial t},
\end{equation}
we find that
\begin{eqnarray}
\biggl ({\partial \over\partial t}+L\biggr )f_{0}&=&-{\beta\over
  C''(f_{0})}\biggl ({\partial\lambda\over\partial t}+{\bf
  u}\cdot\nabla\lambda - \left( \nabla{\bf u} \right)^T {\bf w}\cdot{\bf
  w} \biggr )\nonumber\\
&&-\ {\beta\over C''(f_{0})}\biggl ( {\bf
  w}\cdot\nabla\lambda - \left( \nabla{\bf u} \right)^T {\bf w}\cdot {\bf
  u} - {\bf w}\cdot \nabla\Phi  \biggr ). \label{prec}
\end{eqnarray}
In the above expressions, $\nabla{\bf u}$ designates the matrix $(\partial_{i}u_{j})$ and $T$ is the transposition, i.e. $Aw\cdot u=w\cdot A^{T}u$. Since $\left( \nabla{\bf u} \right)$ is skew-adjoint, i.e. $(\nabla {\bf u})^{T}=-\nabla{\bf u}$, we have $\left(
  \nabla{\bf u} \right) {\bf w}\cdot{\bf w}=0$ and thus,
\begin{equation}
\label{ce14}
(Id - \Pi) \biggl ({\partial \over\partial t}+L\biggr )f_{0} =
  - {\beta\over C''(f_{0})} \biggl ( \nabla\lambda -
\left( \nabla{\bf u} \right) {\bf u} - \nabla\Phi  \biggr )\cdot {\bf
  w}\,.
\end{equation}
Note that the application of $Id-\Pi$ on the first two 
terms of Eq. (\ref{prec}) yields zero since they belong to ${\cal N}$ according to Proposition III.1.  According to Eqs. (\ref{ce17}), (\ref{f1}) and (\ref{ce14}), the
function $f_1$ is determined by the partial differential equation
\begin{equation}
\label{ce20}
C''(f_{0}){\partial\over\partial {\bf v}}\cdot\biggl\lbrack
g(f_{0}){\partial \over\partial {\bf v}} \left( C''(f_0) f_1 \right)
\biggr \rbrack=- \beta  \biggl ( \nabla\lambda -
\left( \nabla{\bf u} \right) {\bf u} - \nabla\Phi  \biggr )\cdot {\bf
  w} + O(\epsilon).
\end{equation}

Introducing the pressure
\begin{equation}
\label{ce21}
p={1\over 3}\int f_{0}w^{2}d^{3}{\bf v},
\end{equation}
and using Eqs. (\ref{ce11}) and (\ref{ce12}), we find after
straightforward calculations that
\begin{equation}
\label{ce22}
{1\over\rho}\nabla p=-\nabla\lambda\,.
\end{equation}
Substituting the foregoing relation in Eq. (\ref{ce20}), we obtain
\begin{equation}
\label{ce25}
C''(f_{0}){\partial\over\partial {\bf v}}\cdot\biggl\lbrack
g(f_{0}){\partial \over\partial {\bf v}} \left( C''(f_0) f_1 \right)
\biggr \rbrack={\beta\over\rho}(\nabla p+\rho\nabla\Phi_{eff})\cdot
{\bf w} +O(\epsilon),
\end{equation}
where
\begin{equation}
\label{ce26}
\Phi_{eff}=\Phi-{u^2\over 2}
\end{equation}
is the {\it effective potential} accounting for inertial
forces.

Since Eq. (\ref{ce25}) is linear in $f_1$, we look for solutions of the form
\begin{equation}
\label{ce27}
C''(f_0) f_1 =-{\beta\over\rho}{\bf R}({\bf w})\cdot(\nabla
p+\rho\nabla\Phi_{eff})+O(\epsilon).
\end{equation}
Substituting this in Eq. (\ref{ce25}), we find that ${\bf R}$ must
satisfy the equation
\begin{equation}
\label{ce28}
C''(f_{0}){\partial\over\partial {\bf w}}\biggl\lbrack
g(f_{0}){\partial R_{i}\over\partial {\bf w}}\biggr \rbrack=-w_{i}.
\end{equation}
In other words, $DQ(f_0)(R_i/C''(f_0)) = -w_i/C''(f_0)$, which yields,
together with (\ref{ce27}),
\begin{equation}
\label{volvic}
-\epsilon\nabla\cdot\biggl (\int f_{1}{\bf v}d^{3}{\bf
  v}\biggr ) = \nabla\cdot\biggl \lbrack \mathcal{X}(\nabla
  p+\rho\nabla\Phi_{eff}) \biggr \rbrack,
\end{equation}
where the matrix $\mathcal{X}=\left( \mathcal{X}_{i,j} \right)_{1\le
  i,j\le 3}$ is given by
\begin{equation}
\label{evian}
\mathcal{X}_{i,j} = - \frac{\epsilon \beta}{\rho}\ \int R_i\ DQ(f_0)\left(
  \frac{R_j}{C''(f_0)} \right)d^{3}{\bf v} =  - \frac{\epsilon
  \beta}{\rho}\ \left\langle \frac{R_i}{C''(f_0)} , DQ(f_0)\left(
    \frac{R_j}{C''(f_0)} \right) \right\rangle_{\mathcal{H}}.
\end{equation}
In particular, $\mathcal{X}$ is a non-negative matrix by
Proposition~\ref{prlin}. Inserting the relation (\ref{volvic}) in
Eq.~(\ref{ce9}), we obtain the generalized Smoluchowski equation
\begin{equation}
\label{ce33}
{\partial\rho\over\partial t}+\nabla\cdot(\rho {\bf u} )=\nabla\biggl
\lbrack \mathcal{X}(\nabla p+\rho\nabla\Phi_{eff})\biggr \rbrack.
\end{equation}

\medskip

Coming back to the equation (\ref{ce28}) satisfied by ${\bf R}$, it
can be written using Eq. (\ref{ce3}) as
\begin{equation}
\label{ce29}
-h(f_{0})\Delta_{w}{\bf R}+g'(f_{0})\beta ({\bf w}\cdot
 \nabla_{w}){\bf R}={\bf w}.
\end{equation}
This equation admits solutions of the form
\begin{equation}
\label{ce30}
{\bf R}({\bf w})={1\over\beta}\phi(x){\bf w},
\end{equation}
where $x=\beta {w^{2}/2}$ and $\phi(x)$ is a solution of the
differential equation
\begin{equation}
\label{ce31}
-h(f_{0})(2 x \phi''+5\phi')+g'(f_{0})(\phi+2 x \phi')=1.
\end{equation}

We have thus established that the first order correction to the
distribution function $f_{0}$ is given by
\begin{equation}
\label{ce32}
f_{1}=-{1\over C''(f_{0})}\phi(x) {1\over\rho} (\nabla
p+\rho\nabla\Phi_{eff})\cdot {\bf w},
\end{equation}
and the matrix $\mathcal{X}$ defined in Eq.~(\ref{evian}) by
\begin{equation}
\label{ce34}
\mathcal{X} = \chi I_3 \;\;\mbox{ with }\;\; \chi={\epsilon\over
  3\rho}\int {1\over C''(f_{0})}\phi(x)w^{2}d^{3}{\bf w}.
\end{equation}

Equations (\ref{ce33}), (\ref{ce31}) and (\ref{ce34}) formally solve
the problem in the general case. We shall now consider simplified
forms of the generalized Kramers equation, corresponding to specific
expressions of $g$ and $h$ \cite{gkt}.  First, we impose $g(f)=f$ and $h(f)=f
C''(f)$. In that case, the collision operator takes the form
\begin{equation}
\label{ce35}
Q(f)={\partial\over\partial {\bf v}}\cdot\biggl\lbrack f C''(f){\partial
  f\over\partial {\bf v}}+\beta f {\bf w}\biggr\rbrack,
\end{equation}
and the differential equation (\ref{ce31}) becomes
\begin{equation}
\label{ce36}
-f_{0}C''(f_{0})(2 x \phi''+5\phi')+\phi+2 x \phi'=1.
\end{equation}
Clearly, $\phi=1$ is a solution. Substituting this result in
Eq. (\ref{ce34}), we obtain
\begin{eqnarray}
\label{ce37}
\chi={\epsilon\over 3\rho}\int {1\over C''(f_{0})}w^{2}d^{3}{\bf
  w}=-{\epsilon\over 3\beta\rho}\int {\partial f_{0}\over\partial {\bf
    w}}\cdot {\bf w}d^{3}{\bf w}\nonumber\\
={\epsilon\over \beta\rho}\int f_{0}d^{3}{\bf
  w}={\epsilon\over\beta}={1\over D\beta}={1\over\xi}.
\end{eqnarray}
Therefore, for this special class of generalized Kramers equations,
the Smoluchowski equation is
\begin{equation}
\label{ce38}
{\partial\rho\over\partial t}+\nabla\cdot( \rho {\bf u} )=\nabla\cdot\biggl
\lbrack {1\over\xi}(\nabla p+\rho\nabla\Phi_{eff})\biggr \rbrack.
\end{equation}
This equation was previously derived in \cite{gfp} by working out
the moments of the Kramers equation. The present method, in addition
of being more rigorous, also provides the first order correction to
the distribution function. We have
\begin{equation}
\label{ce39}
f=f_{0}-\epsilon {1\over C''(f_{0})} {1\over\rho} (\nabla
p+\rho\nabla\Phi_{eff})\cdot {\bf w}+O(\epsilon^{2}).
\end{equation}
If we now impose $g(f)={1/C''(f)}$ and $h(f)=1$, the collision
operator takes the form
\begin{equation}
\label{ce40}
Q(f)={\partial\over\partial {\bf v}}\cdot\biggl\lbrack {\partial
  f\over\partial {\bf v}}+{\beta\over C''(f)} {\bf w}\biggr\rbrack,
\end{equation}
and the differential equation (\ref{ce31}) becomes
\begin{equation}
\label{ce41}
-(2 x \phi''+5\phi')-{C'''(f_{0})\over C''(f_{0})^{2}}(\phi+2 x \phi')=1.
\end{equation}
There does not seem to be any simple solution to this equation. This
implies that $\chi$ is probably not equal to $1/\xi$ in that case.

\section{A moment method for  the generalized Kramers equation}
\label{sec_mom}

In this section, we use a moment method to derive a variant of the 
Smoluchowski equation from the generalized Kramers equation. This
approach was previously developed in \cite{csr} for the fermionic
Kramers equation. We consider here a more general situation. We show in
particular that the diffusion coefficient obtained by the moment
method has a different expression from that obtained by the above
Chapman-Enskog approach.  To simplify the presentation, we consider
the case ${\bf u}=0$, that is
\begin{equation}
\label{m1}
{\partial f\over\partial t}+{\bf v}\cdot{\partial f\over\partial {\bf
    r}}-\nabla\Phi\cdot{\partial f\over\partial {\bf
    v}}={\partial\over\partial {\bf v}}\cdot\biggl\lbrace D\biggl\lbrack
h(f){\partial f\over\partial {\bf v}}+\beta g(f) {\bf
  v }\biggr\rbrack\biggr\rbrace = Q(f)({\bf v}), 
\end{equation}
and seek equations satisfied by the macroscopic quantities
associated with $f$ (mass, impulse and internal energy) defined as
\begin{equation}
\label{m2}
\left(\begin{array}{l}  \rho \\  \rho \ \bar {\bf u} \\ \rho\ \bar e
\end{array} \right) =  \int f({\bf v}) \left( \begin{array}{l}
 1 \\ {\bf v}\\ \displaystyle{{w^{2}\over 2}} \end{array}\right)
 d^3{\bf v}, 
\end{equation}
where ${\bf w}={\bf v}-\bar {\bf u}({\bf r},t)$. Integrating
Eq. (\ref{m1}) against $1$, ${\bf v}$, and $v^2/2$, we
classically get 
\begin{equation}
\label{m3}
\begin{array}{l}
\partial_t  \rho + \nabla\cdot ( \rho \bar {\bf  u})=0,\\
 \\
\partial_t (\rho \bar {\bf u}) + \nabla\cdot ( \rho \bar
 {\bf u} \bar {\bf u}) + \nabla \cdot \bar {P} +   \rho \nabla \Phi=
 \int Q(f) {\bf v} d^3{\bf v},\\  
 \\
\partial_t \bar E + \nabla\cdot \left [\bar E \bar {\bf
 u} + \bar {P} \bar {\bf u} + \bar {\bf q}\right] +  \rho \bar {\bf
 u}\cdot \nabla \Phi = {1\over 2}  \int Q(f) v^2 d^3{\bf v},
\end{array}
\end{equation}
where
\begin{equation}
\label{m4}
\bar E=\rho ({\bar {\bf u}^{2}\over 2}+\bar {e}),\ \ \bar {P} = \int
f({\bf v}) {\bf w}{\bf w} d^{3}{\bf v}, \ \ \mbox{and} \ \ \bar {\bf q} =
\int f({\bf v}) {w^2\over 2} {\bf w} d^3{\bf v},
\end{equation}
are the energy, the pressure tensor and the current of heat. By
combining Eqs. (\ref{m3}-b) and (\ref{m3}-c), we can rewrite the
equation for the energy as  
\begin{equation}
\label{m4bis}
\partial_{t}(\rho \bar {e})+\nabla\cdot (\rho \bar {e} \bar {\bf
  u})+\nabla \cdot \bar {\bf q}+\bar {P}\nabla \bar {\bf u}={1\over
  2}\int Q(f) w^2 d^3{\bf v}.
\end{equation}

System (\ref{m3}) is not closed in the sense that it involves unknown
quantities $\bar{P}(f)$ and $\bar{\bf q}(f)$ that still depend on the
distribution function $f$.  It is well known that a maximum entropy
principle can be used to specify $f$ in terms of $\rho$, $\bar {\bf
u}$ and $\bar {e}$ leading to a closed set of equations for these
quantities.  Under some suitable hypotheses on the entropy functional
$C(f)$, one can show that the distribution function $f_{*}$ that
maximizes the entropy density $-\int C(f) d^3 {\bf v}$ with the
prescribed conditions (\ref{m2}) is given by
\begin{equation}
\label{m5}
C'(f_{*})=-\bar {\beta}({\bf r},t)\biggl ({w^{2}\over 2}+\lambda({\bf
  r},t)\biggr ). 
\end{equation}
Then, we have 
\begin{equation}
\label{m6}
\bar {P}(f_*) = \bar {p} \ I_3,\ \ \bar {p}=\frac{1}{3} \int f_*({\bf
  v}) w^2 d^3{\bf v}, \ \ \mbox{and} \ \ \bar{\bf q}(f_*) ={\bf 0}.
  \end{equation} 
We note in particular that $\bar p={2\over 3}\rho \bar e$. On the
other hand, a straighforward computation gives 
\begin{equation}
\label{m7}
\int Q(f_*) {\bf v} d^3{\bf v} = - \frac{1}{\bar \chi}  \rho \ \bar
{\bf u}, \ \ \  
\end{equation}
and 
\begin{equation}
\label{m7-2}
\int Q(f_{*})  \frac{w^2}{2} d^3{\bf v}=\frac{1}{\epsilon}(\bar \beta
-  \beta)\int w^2 g(f_{*}) d^3{\bf v}, 
\end{equation}
with
\begin{equation}
\label{m8}
\bar \chi  = \epsilon  \frac{\rho}{\beta} \frac{1}{\int g(f_*) d^3{\bf
    v}}. 
\end{equation}
We thus obtain the system of hydrodynamic equations
\begin{equation}
\label{m3b}
\begin{array}{l}
\partial_t  \rho + \nabla\cdot ( \rho \bar {\bf  u})=0,\\
 \rho (\partial_t \bar {\bf u} + \bar  {\bf u}\cdot
 \nabla \bar {\bf u}) =-\nabla \bar p-\rho \nabla
 \Phi-{1\over\bar{\chi}}\rho\bar{\bf u},\\  
 \rho (\partial_{t} \bar e+\bar{\bf u}\cdot \nabla \bar
 e)+\bar{p}\nabla\cdot {\bf u}=\frac{1}{\epsilon}(\bar \beta -
 \beta)\int w^2 g(f_{*}) d^3{\bf v}.
\end{array}
\end{equation}

This system of equations can be simplified in some particular
situations. In the case $g(f)=f$ and $h(f)=fC''(f)$, we find that 
\begin{equation}
\label{m9}
\bar{\chi}={1\over \xi}, \qquad \int w^2 g(f_{*}) d^3{\bf v}=2\rho \bar e.
\end{equation}
In the case $h(f)=1$ and $g(f)=1/C''(f)$, we obtain 
\begin{equation}
\label{m9b}
\bar{\chi}={\rho\over D\beta}{1\over \int {d^{3}{\bf v}\over C''(f)}},
\qquad \int w^2 g(f_{*}) d^3{\bf v}={3\rho\over\bar{\beta}}, 
\end{equation}
where we have used a method similar to that of Eq. (\ref{ce37}) to
obtain the second equality. For the fermionic Kramers equation with
$h(f)=1$ and $g(f)=f(\eta_{0}-f)$, we recover the equations obtained
in \cite{csr}. Finally, for the classical Kramers equation with
$h(f)=1$ and $g(f)=f$, we have $p=\rho\bar{T}$ and
$\bar{e}={3\over 2}\bar{T}$ and the hydrodynamic equations 
\begin{equation}
\label{m3c}
\begin{array}{l}
\partial_t  \rho + \nabla\cdot ( \rho \bar {\bf  u})=0,\\
\rho (\partial_t \bar {\bf u} + \bar  {\bf u}\cdot \nabla \bar {\bf
  u}) =-\nabla (\rho\bar{T})-\rho \nabla \Phi-\xi\rho\bar{\bf u},\\ 
{3\over 2} (\partial_{t}\bar{T}+\bar{\bf u}\cdot \nabla
\bar{T})+\bar{T}\nabla\cdot {\bf u}=3\xi (T -  \bar{T}). 
\end{array}
\end{equation}

Returning to the general case (\ref{m3b}) and considering the strong friction
limit $\epsilon\rightarrow 0$, we get $\rho \bar {\bf u}=
O(\epsilon)$, and the two last  equations  of (\ref{m3b}) give 
\begin{equation}
\label{m9c}
 \rho \bar  {\bf u}  = \bar \chi (\nabla \bar p + \rho \nabla \Phi) +
 O(\epsilon^2), 
\ \ \ \mbox{and} \ \ \ \bar \beta = \beta + O(\epsilon).
\end{equation}
Inserting this expression into the continuity equation (\ref{m3b}-a), we get
\begin{equation}
\label{m10}
\partial_t \bar \rho =  \nabla\cdot \left [ \bar \chi (\nabla \bar p +
  \rho \nabla \Phi)\right] 
+O(\epsilon^2).
\end{equation}
This has a similar form as the model obtained in the above section,
but with new formulae for the quantities $\bar p$ and $\bar \chi$. Now
observing that the difference between the distribution functions $f_0$
used in the Chapman-Enskog expansion and $f_*$ (that maximizes the
entropy) is of the order of $\epsilon$, we get $\bar p = p +
O(\epsilon)$. This means that, in formula (\ref{m10}), if one replaces
$\bar p$ by $p$, then one makes an error of the order $\epsilon ^2$
only.  However, it is not clear whether $\chi = \bar \chi +O(\epsilon)
$ or not. Explicit examples are difficult to construct because of the
implicit dependence of $\chi$ in $\rho$, except for the case $g(f)=f$
where we have $\chi = \bar \chi=1/\xi$.

\section{Langevin particles in interaction}
\label{sec_lr}

\subsection{A non-local Smoluchowski equation} \label{sec_nls}

The previous results remain valid when the force is related to the
density by a relation of the form
\begin{equation}
\label{alet}
\Phi({\bf r},t)=\int \theta({\bf
r}'-{\bf r})\rho({\bf r}',t)d^{3}{\bf r}',
\end{equation}
where $\theta({\bf r}'-{\bf r})$ is a binary potential of interaction
depending only on the absolute distance $\vert {\bf r}' - {\bf r}
\vert$ between the particles. In that case, the potential energy reads
\begin{eqnarray}
W={1\over 2}\int \rho\Phi d^{3}{\bf r},
\label{lr1}
\end{eqnarray}
where the $1/2$ factor guarantees that the contribution of a
particle is not counted twice. With this form of interaction, the
generalized Smoluchowski equation (\ref{ce33}) becomes non-local
and takes the explicit form
\begin{equation}
\label{lr2}
{\partial\rho\over\partial t}+{\bf u}\cdot\nabla\rho=
\nabla\cdot\biggl \lbrace \chi\biggl \lbrack p'(\rho)\nabla\rho+\rho\int
{\partial \theta\over\partial {\bf r}}({\bf r}-{\bf r}')\rho({\bf
  r}',t)d^{3}{\bf r}'+\rho\ {\mb\Omega}{\times} {\bf u}\biggr\rbrack
\biggr\rbrace,
\end{equation}
where ${\bf u}={\mb\Omega}{\times} {\bf r}+{\bf U}$. Equation
(\ref{lr2}) was introduced by Chavanis \cite{gfp} (see an earlier
version in \cite{csr}). It is valid for an arbitrary equation of state
$p(\rho)$ and an arbitrary potential of interaction $\theta({\bf
r}'-{\bf r})$.  When the potential of interaction is long-range and
attractive, this equation exhibits a rich variety of behaviors
associated with canonical phase transitions and blow up
phenomena. Among long-range potentials, the gravitational potential
plays an important role. In this context, Eq. (\ref{lr2}) has been
studied in \cite{crs} for an isothermal equation of state $p=\rho T$,
in
\cite{cs} for a polytropic equation of state $p=K\rho^{\gamma}$ and in
\cite{ribot} for a Fermi-Dirac equation of state. 
In the isothermal case, Eq. (\ref{lr2}) describes a gas of Brownian
particles in interaction (this is the canonical counterpart of a
Hamiltonian system of particles in interaction). In the non-isothermal
case, it describes Langevin particles in interaction displaying
anomalous diffusion. A formal derivation of Eq. (\ref{lr2}) is
proposed in \cite{chavprep} starting from a generalized class of
stochastic processes and using a mean-field approximation. Equations
similar to non-local generalized Fokker-Planck equations also arise in
the context of 2D turbulence and stellar dynamics
\cite{csr,houches}, and to describe the chemotaxis of bacterial populations
\cite{gfp,bedlewo}. As explained in Refs. \cite{gfp,cs}, Eq. (\ref{lr2}) can
be used alternatively as a numerical algorithm to compute nonlinearly
dynamically stable stationary solutions of the Euler-Jeans equations
in astrophysics. This can be particularly interesting in the case of
rotating stars that are not spherically symmetric.

\subsection{The Lyapunov functional} \label{sec_lyap}

An additional interesting feature of the generalized Kramers and
Smoluchowski equations when the potential $\Phi$ is given by Eq.
(\ref{alet}) is the existence of a Lyapunov functional \cite{gfp}. Indeed,
let us write the generalized Kramers equation in the form
\begin{equation}
\label{l1} {d f\over dt}={\partial\over\partial {\bf
v}}\cdot\biggl\lbrace D\biggl\lbrack h(f){\partial f\over\partial
{\bf v}}+\beta g(f) {\bf
  w}\biggr\rbrack\biggr\rbrace\equiv -{\partial {\bf
    J}_{f}\over\partial {\bf v}},
\end{equation}
where $d/dt=\partial/\partial t+L$ is the material derivative and
${\bf J}_{f}$ is the diffusion current. We introduce the
functional
\begin{equation}
\label{l2} F[f]=E[f]-T S[f]-{\mb\Omega}\cdot {\bf L}[f]- {\bf
U}\cdot {\bf P}[f],
\end{equation}
where
\begin{equation}
\label{l3}
E=\int f {v^{2}\over 2}d^{3}{\bf r}d^{3}{\bf v}+ \frac{1}{2}\ \int
\rho \Phi d^{3}{\bf r},
\end{equation}
\begin{equation}
\label{l4}
{\bf L}=\int f {\bf r}{\times} {\bf v} d^{3}{\bf r}d^{3}{\bf v},
\end{equation}
\begin{equation}
\label{l5}
{\bf P}=\int f {\bf v} d^{3}{\bf r}d^{3}{\bf v},
\end{equation}
are the energy, the angular momentum and the impulse respectively.
The ``generalized entropy'' $S$ is defined by Eqs. (\ref{ce2}) and
(\ref{ce3}). The functional (\ref{l2}) can be interpreted as a
generalized free energy. Using the skew-adjointness of
$\nabla{\bf u}$, it is straightforward to check that
\begin{equation}
\label{l6} \dot F=-\int {J_{f}^{2}\over \xi g(f)}d^{3}{\bf
v}d^{3}{\bf r}\le 0,
\end{equation}
so that $F[f]$ plays the role of a Lyapunov functional for the
generalized Kramers equation. A stationary solution of Eq. (\ref{l1}),
defined by $\partial f/\partial t=0$, satisfies $\dot F=0$. Hence, it
cancels the diffusion current ${\bf J}_{f}={\bf 0}$ in virtue of
Eq. (\ref{l6}). It must also cancel the advection term $Lf=0$. Using
these two conditions, we can show that $f_{eq}$ is determined by
\begin{equation}
\label{l7}
C'(f_{eq})=-\beta \biggl ({w^{2}\over 2}+\Phi_{eff}\biggr )-\alpha.
\end{equation}
Therefore, a stationary solution of the Kramers equation
extremizes the free energy (\ref{l2}) at fixed mass, angular
velocity, linear velocity and temperature. It can be shown
furthermore that only {\it minima} of $F$ are linearly stable via
the Kramers equation \cite{gfp}.

The Lyapunov functional associated with the generalized
Smoluchowski equation (\ref{lr2}) can be obtained from Eq.
(\ref{l2}) by replacing $f$ by its leading term $f_{0}$
\cite{grand,gfp}. Thus,
\begin{equation}
\label{l8}
 F[\rho]=E[f_0]-T S[f_0]-{\mb\Omega}\cdot {\bf L}[f_0]- {\bf
U}\cdot {\bf P}[f_0].
\end{equation}
According to Eq. (\ref{ce10}), $f_{0}$ can be written
\begin{equation}
\label{l9}
f_{0}=F\biggl\lbrack \beta\biggl ({w^{2}\over 2}+\lambda({\bf
  r},t)\biggr )\biggr\rbrack,
\end{equation}
where $F(x)=(C')^{-1}(-x)$. Using Eq. (\ref{l9}), the density
$\rho=\int f_{0}d^{3}{\bf v}$
and the pressure $p={1\over 3}\int f_{0} w^{2}d^{3}{\bf v}$ can be put
in the form
\begin{equation}
\label{l10}
\rho={1\over \beta^{3/2}}G(\beta\lambda), \qquad p={1\over
  \beta^{5/2}}H(\beta\lambda),
\end{equation}
where
\begin{equation}
\label{l11} G(x)=4\pi\sqrt{2}\ \int_{0}^{+\infty} F(x+t)t^{1/2}dt,
\end{equation}
\begin{equation}
\label{l12} H(x)={8\pi\sqrt{2}\over 3}\ \int_{0}^{+\infty}
F(x+t)t^{3/2}dt.
\end{equation}
We can now express $E$ and $S$ as functionals of $\rho$. We follow
the derivation given in \cite{grand}. The energy (\ref{l3}) is easily expressed in terms of hydrodynamical variables as 
\begin{equation}
\label{l13}
E[f_{0}]={3\over 2}\int p \ d^{3}{\bf r}+{1\over 2}\int\rho\Phi  d^{3}{\bf
  r}+{1\over 2}\int\rho u^{2} d^{3}{\bf r}.
\end{equation}
On the other hand, the generalized entropy (\ref{ce2}) can be written
\begin{eqnarray}
{S}[f_0]=-{4\pi\sqrt{2}\over\beta^{3/2}}\int d^{3}{\bf
  r}\int_{0}^{+\infty}C\lbrack F(t+\beta\lambda)\rbrack \ t^{1/2}dt.
\label{l15}
\end{eqnarray}
Integrating by parts  and using $C'\lbrack F(x)\rbrack=-x$, we find that
\begin{eqnarray}
{S}[f_0]=-{8\pi\sqrt{2}\over 3\beta^{3/2}}\int d^{3}{\bf r}\int_{0}^{+\infty}
F'(t+\beta\lambda)(t+\beta\lambda)t^{3/2}dt.
\label{l16}
\end{eqnarray}
Integrating by parts one more time and using Eqs. (\ref{l10}),
(\ref{l11}) and (\ref{l12}), we finally obtain
\begin{eqnarray}
{S}[f_0]={5\over 2}\beta\int p \ d^{3}{\bf r}+\beta\int \lambda\rho\ d^{3}{\bf r}.
\label{l17}
\end{eqnarray}
The angular momentum and the linear impulse can be expressed as
\begin{eqnarray}
{\bf L}[f_0]=\int\rho {\bf r}{\times} {\bf u}d^{3}{\bf r},
\label{l18}
\end{eqnarray}
\begin{eqnarray}
{\bf P}[f_0]=\int\rho {\bf u}d^{3}{\bf r}.
\label{l19}
\end{eqnarray}
Collecting all these results, the free energy (\ref{l8}) becomes
\begin{eqnarray}
{F}[\rho]=-\int\rho \biggl (\lambda+{p\over\rho}\biggr )d^{3}{\bf
r} +{1\over 2}\int\rho\Phi d^{3}{\bf r}-{1\over 2}\int\rho u^{2}
d^{3}{\bf r}. \label{l20}
\end{eqnarray}
Finally, using the relation $H'(x)=-G(x)$ obtained from
Eqs. (\ref{l11}) and (\ref{l12}) by a simple integration by parts, it
is easy to check that Eq. (\ref{l10}) implies
\begin{eqnarray}
\lambda+{p\over\rho}=-\int_{0}^{\rho}{p(\rho')\over\rho^{'2}}d\rho'.
\label{l21}
\end{eqnarray}
Hence, the free energy can be written more explicitly
\begin{eqnarray}
{F}[\rho]=\int \rho \int_{0}^{\rho}{p(\rho')\over\rho^{'2}}d\rho'
d^3{\bf r}+{1\over
  2}\int\rho\Phi d^{3}{\bf r}-{1\over 2}\int\rho u^{2}
d^{3}{\bf r}.
\label{l22}
\end{eqnarray}
This  is the Lyapunov functional for the generalized
Smoluchowski equation (\ref{lr2}). Recalling that
$\chi=\chi(\rho)$ is non-negative, a straightforward calculation
shows that
\begin{eqnarray}
\dot F=-\int {1\over\rho}\ \chi(\rho) {\bf J}_{\rho}^2 d^{3}{\bf
  r}\le 0,
\label{l23}
\end{eqnarray}
where ${\bf J}_{\rho}$ is the diffusion current
\begin{eqnarray}
{\bf J}_{\rho}=-(\nabla p+\rho\nabla\Phi_{eff}).
\label{l24}
\end{eqnarray}
At equilibrium, ${\bf J}_{\rho}={\bf 0}$, and we obtain the
condition of hydrostatic equilibrium in the rotating frame
\begin{eqnarray}
\nabla p+\rho\nabla\Phi_{eff}={\bf 0}. \label{l25}
\end{eqnarray}
\medskip

\subsection{Generalized Cahn-Hilliard equation} \label{sec_ch}

The free energy functional (\ref{l22}) can be written explicitly
\begin{equation}
\label{bordeaux} {F}[\rho]=\int \Gamma(\rho)d^3{\bf r}+{1\over
  2}\int \int \rho({\bf r})\theta({\bf r'} - {\bf r})\rho({\bf r'})
  d^3{\bf r} d^3{\bf r'} -{1\over 2}\int\rho u^{2}
d^{3}{\bf r},
\end{equation}
where we have defined
\begin{equation}
\label{lune}
\Gamma(\rho)=\rho\int_{0}^{\rho}{p(\rho')\over\rho^{'2}}d\rho'.
\end{equation}
Noting that
\begin{equation}
\label{agen} \rho\Gamma''(\rho)=p'(\rho),
\end{equation}
and taking the functional derivative of $F[\rho]$, we observe that
the generalized Smoluchowski equation (\ref{lr2}) can be written
\begin{equation}
\label{toulouse} {\partial\rho\over\partial t} + {\bf
u}\cdot\nabla\rho = \nabla\cdot \biggl( \chi\rho\nabla{\delta
F\over\delta\rho} \biggr)\,.
\end{equation}
If we now consider the case of short-range interactions, it is
possible to expand the potential
\begin{equation}
\label{fes}
\Phi({\bf r},t) = \int \theta({\bf r}')\rho({\bf r}+{\bf r'}) d^3{\bf
  r}'
\end{equation}
in Taylor series for ${\bf r}'\to {\bf 0}$. Introducing the notations
\begin{equation}
\label{marrakech}
a=4\pi \int_0^{+\infty} \theta(x)x^2 dx \;\;\mbox{ and }\;\; b=
{4\pi\over 3} \int_0^{+\infty} \theta(x)x^4 dx\,,
\end{equation}
we obtain to second order
\begin{equation}
\label{ouarzazate}
\Phi({\bf r},t) = a\rho({\bf r},t) + {b\over 2} \Delta\rho({\bf
  r},t)\,.
\end{equation}
In that limit, the free energy takes the form
\begin{equation}
\label{zagoura}
F[\rho] = -{b\over 2} \int \left\{ {(\nabla\rho)^2\over 2} +
  V(\rho) \right\} d^3{\bf r}\,,
\end{equation}
where we have set $V(\rho) = -2\Gamma(\rho)/b - (a/b)\rho^2 +
(1/b)\rho u^2$.  This is the usual expression of the Landau free
energy. In general $b$ is negative so we have to minimize the
functional integral. On the other hand, in this case of short-range
interactions, the conservative equation (\ref{toulouse}) becomes
\begin{equation}
\label{dunkerque}
{\partial\rho\over\partial t} + {\bf u}\cdot\nabla\rho = \nabla\cdot
\biggl\lbrace {b\chi\over 2}\rho\nabla \left( \Delta\rho - V'(\rho) \right)
\biggr\rbrace\,.
\end{equation}
This is the Cahn-Hilliard equation which has been extensively studied
in the theory of phase ordering kinetics. Its stationary solutions
describe ``domain walls''. We can view therefore Eq.~(\ref{lr2}) as a
generalization of the Cahn-Hilliard equation to the case of long-range
interactions and arbitrary free energy functionals. This is why its
physical and mathematical richness is so important. 

\section{Conclusion}
\label{sec_conclusion}

In this paper we have derived the Smoluchowski equation from the
Kramers equation by using a formal Chapman-Enskog expansion. We have
shown furthermore that this derivation could be generalized to a
larger class of Kramers equations where the diffusion coefficient and
the friction term depend on the distribution function. The resulting
Smoluchowski equation takes a nice form where the usual term $T\rho$
is replaced by a pressure $p(\rho)$, as previously noted in
\cite{csr,gfp}.  When the potential $\Phi({\bf r},t)$ is induced by
the density of particles themselves, the generalized Smoluchowski
equation becomes non-local and can generate a rich variety of
phase transitions as discussed in \cite{gfp}, and more
specifically in
\cite{crs,cs,ribot}. In the limit of short-range interactions, it
reduces to the Cahn-Hilliard equation widely studied in the context of
phase ordering kinetics. It is interesting to note that these
different equations are closely related to each other and that they
are associated with an {\it effective} generalized thermodynamical formalism
\cite{gfp,gkt}. An interest of this formalism is to unify different
types of approaches. For example, it encompasses the case of quantum
particles with exclusion or inclusion principle. On the other hand, our
approach extends the nonlinear Fokker-Planck equations \cite{plastino}
associated with the Tsallis entropy. Indeed, the Tsallis entropy does
not play any special role in our formalism and most of the properties
of the nonlinear Fokker-Planck equation can be generalized to a wider
class of entropy functionals \cite{gfp}. The Tsallis entropy forms,
however, an important class of functionals associated with polytropic
distributions and power-laws. These distributions generate a natural
form of self-confinement that can be of interest in nonextensive
systems. However, it was our interest here to show that a more general
formalism could be developed consistently. A notion of ``generalized
thermodynamics'', with a different presentation and a different
motivation, has been developed independently by Kaniadakis
\cite{kaniadakis}, Frank \cite{frank} and Naudts \cite{naudts}.

\acknowledgments

One of us (P.H.C) acknowledges stimulating discussions with
T.D. Frank, G. Kaniadakis, J. Naudts, A. Plastino, P. Quarati and
C. Tsallis during the Next 2003 meeting in Cagliari. Partial support
from the EU network HYKE contract number: HPRN-CT-2002-00282 is
gratefully acknowledged.

\newpage
\appendix

\section{BGK operator}
\label{sec_bgk}

In this Appendix, we derive the generalized Smoluchowski equation from
a simplified form of kinetic equations. Specifically, we consider a
BGK collision operator so that
\begin{equation}
\label{b1}
{\partial f\over\partial t}+{\bf v}\cdot{\partial f\over\partial {\bf
    r}}-\nabla\Phi\cdot{\partial f\over\partial {\bf
    v}}=-{1\over\epsilon}{f-f_{0}\over\tau},
\end{equation}
where $\epsilon\tau=\epsilon\tau(\rho)$ is a ``typical collision''
time and $f_{0}$ is defined by
\begin{equation}
\label{b2}
C'(f_{0})=-\beta \biggl ({w^{2}\over 2}+\lambda({\bf r},t)\biggr ),
\end{equation}
with $\int f_{0}d^{3}{\bf v}=\rho$, ${\bf u}={\bf U}+{\bf
  \Omega}{\times}{\bf r}$ and ${\bf w} = {\bf v} - {\bf u}$. We note
that contrary to the usual BGK operator, the velocity distribution
  $f_{0}$ is {\it isotropic} (in the rotating frame). This is
  consistent with the properties of the generalized Kramers
  equation (\ref{ce1}). We consider the limit $\epsilon\rightarrow 0$. Writing
  $f_1=\left( f-f_{0} \right)/\epsilon$, we get to leading order
\begin{equation}
\label{b3}
f_1=-\tau\biggl ({\partial\over\partial t}+L\biggr )f_{0} + O(\epsilon).
\end{equation}
Integrating Eq. (\ref{b1}) over the velocities, we obtain
\begin{equation}
\label{b4}
{\partial\rho\over\partial t}+\nabla\cdot\biggl (\int f {\bf v}d^{3}{\bf
  v}\biggr )=0,
\end{equation}
or, using Eq. (\ref{b2}),
\begin{equation}
\label{b5}
{\partial\rho\over\partial t}+\nabla\cdot( \rho {\bf u})
=-\epsilon\nabla\cdot\biggl( \int f_1 {\bf v}d^{3}{\bf v}\biggr ).
\end{equation}
Now,
\begin{equation}
\label{b6}
\int f_1 {\bf v}d^{3}{\bf v}=-\tau\int {\bf v} \biggl
({\partial\over\partial t}+L\biggr )f_{0}d^{3}{\bf v} + O(\epsilon).
\end{equation}
The first term in the foregoing equation can be written
\begin{equation}
\label{b7}
\int {\bf v} {\partial f_{0}\over\partial t} d^{3}{\bf
  v}={\partial\over\partial t}\int f_{0}{\bf v} d^{3}{\bf v}={\bf u}
  {\partial\rho\over\partial t}.
\end{equation}
To leading order, Eq. (\ref{b5}) reduces to
\begin{equation}
\label{b8}
{\partial\rho\over\partial t}+\nabla\cdot(\rho {\bf u})=O(\epsilon).
\end{equation}
Therefore,
\begin{equation}
\label{b9}
\int {\bf v} {\partial f_{0}\over\partial t} d^{3}{\bf
  v}=-{\bf u}\nabla\cdot(\rho {\bf u})+O(\epsilon).
\end{equation}
The second term in Eq. (\ref{b6}) can be written
\begin{equation}
\label{b10}
\int v_i L f_{0} d^{3}{\bf v}= \int v_{i}\biggl (v_{j}{\partial
  f_{0}\over \partial x_{j}}-{\partial\Phi\over\partial
  x_{j}}{\partial f_{0}\over\partial v_{j}}\biggr )d^{3}{\bf
  v}={\partial \over\partial x_{j}}\int f_{0}v_{i}v_{j}d^{3}{\bf
  v}+\rho{\partial\Phi\over\partial x_{i}}.
\end{equation}
Now,
\begin{equation}
\label{b11}
\int f_{0}v_{i}v_{j}d^{3}{\bf v}=p\delta_{ij}+\rho u_i u_j,
\end{equation}
where $p={1\over 3}\int f_{0}w^{2}d^{3}{\bf v}$ is the pressure. After
straightforward calculations, we find that
\begin{equation}
\label{b12}
{\partial \over\partial x_{j}}\int f_{0}v_{i}v_{j}d^{3}{\bf
  v}={\partial p\over \partial x_i} + u_i \nabla\cdot( \rho {\bf u} )+\rho
  \left( {\mb\Omega}{\times} {\bf u} \right)_i.
\end{equation}
Combining the foregoing results, we obtain
\begin{equation}
\label{b13}
\int f_1 {\bf v}d^{3}{\bf v}=-\tau (\nabla p+\rho\nabla\Phi+\rho
{\mb\Omega}{\times} {\bf u})+O(\epsilon).
\end{equation}
Substituting this in Eq. (\ref{b5}) and omitting terms of order
$\epsilon^2$, we obtain the generalized Smoluchowski equation
\begin{equation}
\label{b14}
{\partial\rho\over\partial t}+\nabla\cdot( \rho {\bf u}
)=\epsilon\nabla\cdot\biggl \lbrack \tau (\nabla
p+\rho\nabla\Phi_{eff})\biggr \rbrack.
\end{equation}

\section{Time dependent Lagrange multipliers}
\label{sec_time}

In this Appendix, we consider the generalized Kramers equation
\begin{equation}
\label{t1}
{\partial f\over \partial t}+Lf={\partial\over\partial {\bf
    v}}\cdot\biggl\lbrace D\biggl\lbrack h(f){\partial f\over\partial
  {\bf v}}+\beta(t) g(f) ({\bf v}-{\mb\Omega}(t){\times} {\bf r}-{\bf
  U}(t))\biggr\rbrack\biggr\rbrace,
\end{equation}
where the parameters $\beta$, ${\mb\Omega}$ and ${\bf U}$ depend on
time so as to satisfy at each instant $t$, the conservation of energy,
angular momentum and impulse, and the potential $\Phi$ is given by
Eq. (\ref{alet}). Their evolution is obtained by
substituting Eq. (\ref{t1}) in the constraints $\dot E=\dot {\bf
L}=\dot {\bf P}={\bf 0}$. It can be shown that Eq. (\ref{t1})
increases the generalized entropy (\ref{ce2}) at fixed mass,
energy, angular momentum and impulse (H-theorem). This corresponds to a
microcanonical situation. Equation (\ref{t1}) can be
obtained from a variational principle called Maximum Entropy
Production Principle (MEPP). In this context, $\beta$, ${\mb\Omega}$
and ${\bf U}$ appear as time dependant Lagrange multipliers associated
with the integral constraints. This type of equations were proposed in
\cite{csr} as a small-scale parametrization of the gravitational
Vlasov-Poisson system in the context of the theory of violent
relaxation. As explained in \cite{gfp}, these relaxation equations can
also be used as numerical algorithms to compute nonlinearly
dynamically stable stationary solutions of the Vlasov-Poisson
system. This would be particularly interesting in the case of rotating
stellar systems that are non spherically symmetric. They can have
similar applications in other domains of physics with different types
of interaction.

Considering the limit $\xi\rightarrow +\infty$, the Chapman-Enskog
method of Sec.~\ref{sec_ce} can be implemented with only slight
modifications. Due to the time dependence of the Lagrange multipliers,
Eq. (\ref{ce13}) is replaced by
\begin{equation}
\label{t2}
C''(f_{0}){\partial f_{0}\over\partial
  t}=-\beta{\partial\lambda\over\partial t}+ \beta {\bf
  w}\cdot\dot{\bf u}-\dot\beta \biggl ({w^{2}\over 2}+\lambda \biggr ),
\end{equation}
where $f_{0}$ is still given by Eq. (\ref{ce10}) and ${\bf u} = {\bf
  U}+{\bf \Omega}{\times} {\bf r}$. Consequently, introducing the notation
\begin{equation}
\label{t4}
\nu=\int {1\over C''(f_{0})}d^{3}{\bf v},
\end{equation}
and using the identity
\begin{equation}
\label{t5}
\int {1\over C''(f_{0})}{w^{2}\over 2}d^{3}{\bf v}=-{1\over
  2\beta}\int {\partial f_{0}\over\partial {\bf w}}{\bf w}d^{3}{\bf
  v}={3\rho\over 2\beta},
\end{equation}
resulting from Eq. (\ref{ce11}),
Eq.~(\ref{ce14}) becomes
\begin{equation}
\label{t3}
(Id - \Pi) \biggl ({\partial \over\partial t}+L\biggr )f_{0} =
  - {\beta\over C''(f_{0})} \biggl ( \nabla\lambda - \dot{\bf u} -
\left( \nabla{\bf u} \right) {\bf u} - \nabla\Phi  \biggr )\cdot {\bf
  w} - {\dot\beta\over C''(f_0)} \biggl ( {w^2\over 2} - {3\rho\over
  2\nu\beta} \biggr)\,.
\end{equation}
Using Eqs. (\ref{t3}), we find that Eq. (\ref{ce20}) is replaced by
\begin{eqnarray}
\nonumber
f_1 & = & -\beta\ DQ(f_0)^{-1}  \biggl ( {1\over C''(f_{0})} \left(
  \nabla\lambda - \dot{\bf u} - \left( \nabla{\bf u} \right) {\bf u} -
  \nabla\Phi  \right)\cdot {\bf w} \biggr )\\
\label{t11}
& & - \ \dot\beta\ DQ(f_0)^{-1}  \biggl ( {1\over C''(f_0)} \left(
  {w^2\over 2} - {3\rho\over 2\nu\beta} \right) \biggr ).
\end{eqnarray}
Since $f_{0}({\bf w})$ is spherically symmetric, only the terms
proportional to ${\bf w}$ in the right-hand side of Eq. (\ref{t11})
will give a non-zero contribution in Eq. (\ref{ce9}). Therefore,
Eq. (\ref{ce33}) remains valid with the substitution
$\nabla\Phi\rightarrow\nabla\Phi+\dot{\bf u}$. We thus obtain the
generalized Smoluchowski equation
\begin{equation}
\label{t12}
{\partial\rho\over\partial t}+\nabla\cdot ( \rho {\bf u}
)=\nabla\cdot\biggl \lbrack\chi(\nabla p+\rho\nabla\Phi_{eff}+\rho
\dot{\bf u})\biggr \rbrack.
\end{equation}
Note that the last term was forgotten in \cite{csr}. The time
dependent Lagrange multipliers $\beta(t)$, ${\bf \Omega}(t)$ and ${\bf
  U}(t)$ are determined by replacing $f$ by its leading term $f_0$ in
the conservation of energy, angular momentum and impulse. Recalling
(\ref{l13}), (\ref{l18}) and (\ref{l19}), we obtain
\begin{eqnarray*}
{3\over 2}\int p(\rho) d^{3}{\bf r}+{1\over 2}\int\rho\Phi
  d^{3}{\bf r}+{1\over 2}\int\rho u^{2}d^{3}{\bf r} & = & \mbox{
  const.}\,,\\
\int \rho {\bf r}{\times}{\bf u}d^3{\bf r} & = & \mbox{const.}\,,\\
\int \rho {\bf u}d^3{\bf r} & = & \mbox{const.}\,,
\end{eqnarray*}
where $p(\rho)$ is given by (\ref{l10}), (\ref{l11}) and (\ref{l12}).
This type of equations with time dependant temperature has been
studied in \cite{crs,cs,rosier,biler,guerra}.

\end{document}